**A senescent-immune reserve niche model for incomplete lobular involution in the aging breast**

Jaida C. Lue[1], Darren J. Baker[2], Amy C. Degnim[3], Stacey J. Winham[4], Mark E. Sherman[5], Derek C. Radisky[1]

[1]Department of Cancer Biology, Mayo Clinic, Jacksonville, FL, USA
[2]Departments of Biochemistry and Molecular Biology and Pediatrics, Mayo Clinic, Rochester, MN, USA
[3]Department of Surgery, Mayo Clinic, Rochester, MN, USA
[4]Department of Quantitative Health Sciences, Mayo Clinic, Rochester, MN, USA
[5]Department of Quantitative Health Sciences, Mayo Clinic, Jacksonville, FL, USA

## Abstract
Breast cancer incidence rises with age and peaks across the menopausal transition, yet why some postmenopausal lobules persist, and why that persistence predicts cancer risk, remains unresolved. Incomplete age-related lobular involution is one of the strongest tissue-level predictors of subsequent breast cancer, but it is still commonly viewed as passive failure of hormonally driven regression. This Review proposes a different framework: persistent lobules are maintained by an active reserve niche that outlasts its reproductive function. Recent human tissue studies show that persistent postmenopausal lobules retain a proliferation-competent epithelial state and that immune cells accumulate at the lobular boundary without being positioned to engage p16INK4a-positive epithelial foci; in a separate longitudinal cohort, senescence-linked programs reverse their association with involution pace across menopause. We integrate these findings with breast epidemiology and mammary stromal biology, and draw on reserve systems in skeletal muscle, hematopoiesis, and postmenopausal endometrium as sources of candidate mechanisms and testable predictions rather than as evidence for the breast. We argue that menopause is a biological control point at which tissue fate diverges, and that what distinguishes persistence from resolution is not the amount of inflammation present but whether immune effectors are spatially positioned to execute clearance. This shifts attention from epithelial quantity to microenvironmental state and identifies the perimenopausal window as a promising interval for biomarker-guided risk stratification and prevention.

## Introduction
Breast cancer is largely a disease of aging: incidence rises steeply with age, the median age at diagnosis is approximately 62 years, and roughly 84% of cases occur in women over 50 (Giaquinto, Sung et al. 2024); age alone explains more of the incidence curve than any other single variable, and approximately half of all breast cancers occur in women without identifiable genetic or lifestyle risk factors (Lue and Radisky 2025). Age-related lobular involution (ARLI) is the defining structural aging event of the breast: the progressive regression of terminal duct lobular units (TDLUs), the primary functional and cancer-susceptible epithelial structures of the gland. This process begins in the fourth decade, accelerates around the menopausal transition, and continues variably for decades into postmenopause (Figueroa, Pfeiffer et al. 2014, Radisky, Visscher et al. 2016). Its completion carries direct clinical significance. Postmenopausal women with incomplete ARLI face up to a threefold elevated risk of subsequent breast cancer compared to women whose lobular regression was complete, independently of parity, body mass index, and family history (Milanese, Hartmann et al. 2006), and stagnation of the involution process, the failure to progress toward completion, is itself a significant predictor of elevated risk (Radisky, Visscher et al. 2016).

What has resisted explanation is why. The dominant interpretation is essentially quantitative: more residual epithelial tissue means more target cells. But this framing has important limitations. Breast cancer risk does not scale continuously with residual lobular density; instead it bifurcates around a completion threshold, with partial and absent involution conferring substantially greater risk than even modest degrees of completion (Milanese, Hartmann et al. 2006). Involution status is independent of other markers of breast density and epithelial content, suggesting it carries biological information beyond tissue quantity (Milanese, Hartmann et al. 2006, Radisky, Visscher et al. 2016). Most directly: local inflammatory proteins in histologically normal breast tissue from breast cancer patients are inversely associated with complete involution, with higher levels of pro-inflammatory markers IL-6, TNF-α, CRP, COX-2, leptin, SAA1, and IL-8, and notably the immunosuppressive cytokine IL-10, all correlating with less complete lobular regression after age adjustment (Hanna, Dumas et al. 2017). The microenvironment of persistent lobules is biologically active, not neutral. The more fundamental question is therefore not what prevents involution from occurring, but what actively maintains the tissue that fails to involute.

This review proposes that incomplete ARLI represents the persistence of an adaptive reserve tissue program, an actively maintained, locally sustained epithelial state supported by microenvironmental signaling, rather than a passive default arising from incomplete hormonal withdrawal. A conceptual clarification is necessary from the outset: we use "adaptive" in its evolutionary rather than its functional sense; the program that maintains residual lobular tissue was shaped by selection during reproductive years, when episodic involution and potential late pregnancy created genuine biological value for keeping some lobular capacity in reserve. Menopause represents a context this program was never designed to resolve; the maintenance machinery persists, but the hormonal resolution signal that would normally terminate it does not arrive. What is adaptive in origin can therefore be maladaptive in consequence, and understanding that distinction is central to understanding both the biology of incomplete ARLI and the timing of its cancer risk. We develop this argument in four steps: establishing the conceptual framework of adaptive reserve programs by cross-tissue comparison; applying that framework to ARLI using the published breast biology literature; proposing the senescent-immune niche as the mechanistic basis of reserve maintenance in the postmenopausal breast; and identifying the menopausal transition as the control point that determines whether the niche resolves or locks.

A note on how evidence is treated. The argument below draws on three kinds of material, and we have tried to keep them visibly distinct. Some claims are established in human breast tissue and are presented as such. Some are inferred from mouse mammary gland, from cultured human mammary cells, or from reserve systems in other organs; these are presented as sources of candidate mechanisms and testable predictions, not as evidence about the human breast. Some are proposed and untested, and are marked as proposals. Where a claim rests on rodent data we say so in the sentence that makes it, because the species is part of the claim rather than a detail of its citation.

## Reserve Tissue Programs: Conceptual Framework

***Definition and core properties.*** Adult tissues that must sustain regenerative capacity across decades face a fundamental problem: the cells that supply regeneration cannot divide continuously without exhausting themselves or accumulating transformative mutations, yet they must remain available on timescales longer than any single remodeling event. Reserve tissue programs are the biological solution. Operationally, a reserve tissue program maintains a latent regenerative module in a poised state, supported by specialized niches, immune-stromal signaling, and intrinsic molecular restraints, in a manner that enables recovery when conditions improve, while potentially increasing long-term pathology risk if the program remains chronically engaged (de Morree and Rando 2023). This definition distinguishes reserve programs from two superficially similar states: passive quiescence, in which cells simply fail to receive proliferative signals, and terminal differentiation, in which cells have permanently exited the regenerative pool. Reserve programs are actively maintained: their constituents remain metabolically engaged, stress-resistant, and primed for deployment. The apparent stillness of the reserve conceals ongoing molecular activity: suppression of mRNA translation through post-transcriptional mechanisms, metabolic rewiring toward oxidative phosphorylation that reduces DNA damage accumulation, whose purpose is readiness rather than rest (Tumpel and Rudolph 2019, de Morree and Rando 2023).

Reserve programs operate across a spectrum of quiescence depth, from shallow states in which cells are one or two signals from re-entering the cell cycle to deep states requiring more substantial reprogramming for reactivation. Under chronic stress or loss of niche support, reserve cells progress toward the deep end of this spectrum, which merges with irreversible senescence-like arrested states from which reactivation is no longer possible. This quiescence-to-senescence continuum is critical to the ARLI hypothesis: the p16INK4a/Rb axis that enforces poised quiescence can become permanently engaged when the signals maintaining its reversibility are withdrawn or overwhelmed. Incomplete ARLI may initially represent a shallow reserve state, but its trajectory under chronic aging and immune remodeling is toward deeper entrenchment and eventual senescent conversion (Tumpel and Rudolph 2019, de Morree and Rando 2023).

***The resilience–risk tradeoff as organizing principle.*** Reserve tissue programs are Janus-faced. The persistence and molecular poising that enable a tissue to survive chronic stress and regenerate after injury are the same properties that create long-term pathology risk. Long-lived progenitor pools accumulate DNA damage through oxidative stress, mitochondrial dysfunction, and epigenetic drift even without dividing (Tumpel and Rudolph 2019). Their maintenance in a proliferation-competent poised state means that the cellular machinery required for transformation, including cell cycle progression, apoptosis evasion, growth factor responsiveness, is already partially engaged. None of these risks require prior neoplastic conversion of the reserve cells; they are consequences of reserve architecture rather than defects within it.

This tradeoff is the conceptual key to understanding why incomplete ARLI elevates breast cancer risk without requiring persistent lobules to be pre-malignant at the time the epidemiologic association is measured. The elevated risk is a property of the reserve state itself, a maintained, proliferation-competent tissue unit with a chronic inflammatory microenvironment, rather than evidence of transformation already underway. Aging can drive reserve programs toward the risk side through two distinct failure modes: by depleting the reserve too rapidly, so the progenitor pool is exhausted before malignancy can be selected for; or by failing to resolve a reserve that was shaped by selection for a prior biological context, allowing it to become a chronic substrate for pathology once that context no longer obtains. In the aging breast it is the second failure mode that operates: a program selected during reproductive years, now running without the resolution signal that menopause was never reliably shaped to provide, and the comparator systems below illuminate both why that is and what it predicts mechanistically.

***Cross-tissue comparators.*** *Skeletal muscle satellite cells.* Muscle satellite cells are the canonical quiescent reserve stem cell population of adult mammalian tissues. These PAX7-positive cells reside between the plasma membrane and basal lamina of individual myofibers, maintained in G0 by niche-derived signals, including Notch ligands, extracellular matrix cues, and paracrine inputs from vasculature, that enforce quiescence while preserving reactivation competence (de Morree and Rando 2023). In young individuals, satellite cells respond to injury with rapid activation, asymmetric division, and fiber fusion, followed by return to quiescence after repair (**Fig. 1, Zone 1**).

With aging, this cycle breaks down through a mechanism defined with unusual precision in mice. Resting satellite cells progressively lose active suppression of the $p16^{INK4a}$/Rb axis, transitioning into an irreversible pre-senescent state even without injury. When challenged, they undergo geroconversion, accelerated entry into full senescence, even after transplantation into a young host, demonstrating cell-intrinsic failure. In that murine system, silencing $p16^{INK4a}$ is sufficient to restore reversible quiescence and regenerative function, establishing quiescence-to-senescence conversion as the causal mechanism of the regenerative deficit of aged muscle (Sousa-Victor, Gutarra et al. 2014) (**Fig. 1, Zone 2**). The breast reserve program is proposed to follow the same spectrum in the opposite direction: whereas aged murine muscle loses its reserve through senescent conversion, incomplete ARLI retains a reserve maintained in an increasingly senescence-prone state (**Fig. 1, Zone 3**).

*Hematopoietic stem cells.* Hematopoietic stem cells illustrate the risk consequence of long-lived reserve maintenance. HSCs are maintained in stable quiescence within the endosteal niche across decades; unlike satellite cells, aging does not drive senescent conversion, and the reserve persists morphologically unchanged (**Fig. 2, Zone 1**), and the compartment persists as a long-lived progenitor system into later life (Tumpel and Rudolph 2019) (**Fig. 2, Zone 2**). Risk accumulates instead through somatic evolution within the maintained pool. Large population sequencing studies showed that age-associated clonal hematopoiesis becomes increasingly detectable in individuals without overt hematologic disease, and that these clones are associated with increased subsequent hematologic cancer risk (Genovese, Kahler et al. 2014, Jaiswal, Fontanillas et al. 2014, Steensma, Bejar et al. 2015). The key principle is that risk is architectural rather than cellular: because the reserve is maintained across decades, rare advantageous mutations have an extended opportunity for selection and clonal propagation, independent of any pre-malignant transformation in the reserve cells themselves (**Fig. 2, Zone 3**).

*Postmenopausal endometrium.* The postmenopausal endometrium is the closest hormonal analog to the breast in the reserve program framework. Basalis stem and progenitor cells that drive cyclic endometrial regeneration persist in quiescence after menopause, without hormonal stimulation or regenerative demand, maintained for years to decades in atrophic but architecturally intact tissue (Cousins, Filby et al. 2021) (**Fig. 3, Zone 1**). The functional competence of this dormant reserve is demonstrated clinically: exogenous estrogen administered to women with decades of postmenopausal atrophy rapidly restores pre-menopausal endometrial thickness, establishing that reactivation capacity is fully retained regardless of time elapsed since menopause (Cousins, Filby et al. 2021).

Risk arises not from transformation during dormancy but from aberrant reactivation. Unopposed estrogen exposure — from obesity-associated aromatization or exogenous hormone regimens — triggers abnormal proliferation from the maintained reserve, driving endometrial hyperplasia and carcinoma (Nees, Heublein et al. 2022) (**Fig. 3, Zone 2**). The endometrium thus illustrates passive reserve dormancy: the progenitor pool awaits an external hormonal signal, and pathology requires that signal to arrive in unbalanced form. This contrast with the breast is the conceptual hinge of the ARLI hypothesis. In the postmenopausal breast, the equivalent failure is not aberrant reactivation by an external signal but paracrine maintenance without hormonal input; the reserve would, under the reserve niche model proposed in Section 4, be actively sustained by a senescent-immune niche rather than resolved by it (**Fig. 3, Zone 3**).

***What distinguishes ARLI from canonical reserve programs and what it shares.*** Reserve maintenance requires niche integrity, cell-intrinsic quiescence programs, and active regulation along the quiescence-to-senescence continuum (Tumpel and Rudolph 2019, de Morree and Rando 2023). The failure modes differ instructively: in aged muscle, the reserve is depleted through senescent conversion of satellite cells (**Fig. 1**); in the hematopoietic system, the stably maintained reserve becomes a substrate for somatic evolution and clonal malignant risk (**Fig. 2**); in the postmenopausal endometrium, the reserve persists in passive dormancy awaiting an external reactivation signal (**Fig. 3**). In each case, niche integrity is either lost or passively held.

In the postmenopausal breast, the pathology runs in the opposite direction: niche integrity is aberrantly sustained, and that maintenance is itself the problem (**Fig. 4**).

Incomplete ARLI shares the core reserve logic but differs from all three comparators in one architecturally decisive respect (**Fig. 4**). In every canonical reserve system, the unit of reserve is individual cells or small subpopulations within otherwise differentiated tissue. In incomplete ARLI, the reserve unit is an entire functional tissue structure: a TDLU with organized acinar architecture, specialized intralobular stroma, and a spatially distinct paracrine signaling environment. Sustaining it requires not merely keeping cells in a particular molecular state but maintaining the organized multicellular assembly of epithelium, stroma, vasculature, and immune constituents together, without systemic hormonal support. This architectural scale implies a requirement absent from all cell-level reserve systems: the local paracrine environment within and around the TDLU must itself actively generate the maintenance signals that estrogen withdrawal has removed from the systemic circulation.

## ARLI as Reserve Tissue: Evidence from the Published Literature

Having established the reserve program framework through three comparator systems, we now apply it to ARLI directly. The defining question is whether persistent lobular tissue is passively maintained, cells that simply failed to regress, or actively sustained by microenvironmental signaling in the absence of systemic hormonal input. Two converging lines of evidence favor the latter.

Before applying the framework, it is worth being explicit that the three systems above are analogies, and that analogies do not transfer evidence. Skeletal muscle, hematopoiesis, and endometrium establish that actively maintained reserve programs exist, that they are niche-dependent, and that their failure modes differ; they establish nothing about the breast. The relevant question is what has been shown in human breast tissue, and the answer is that some components of the reserve claim are supported and others are not.

Three components are supported. Human breast contains quiescent stem-like cells: keratin-defined subsets of human breast epithelium include a candidate stem cell population described as essentially quiescent, distinguishable from more actively dividing lobular progenitors (Villadsen, Fridriksdottir et al. 2007), and clonal analysis using endogenous mitochondrial marks demonstrates long-lived, multi-lineage stem cells resident in the luminal layer of normal human breast (Cereser, Jansen et al. 2018). Human breast epithelium retains regenerative competence: it forms organoids preserving basal and luminal lineages including estrogen receptor-positive cells (Rosenbluth, Schackmann et al. 2020), and defined human mammary repopulating units regenerate bilayered epithelial structures in vivo (Eirew, Stingl et al. 2008). Progenitor content persists into late life: primary human mammary epithelial cells from donors spanning the second through tenth decades remain culturable and clonogenic, although multipotent progenitors accumulate with a basal differentiation bias and lineage fidelity is progressively lost with age (Garbe, Pepin et al. 2012, Pelissier Vatter, Schapiro et al. 2018, Sayaman, Miyano et al. 2024).

Two components are not supported, and one published observation runs against a straightforward reading of the model. No study has measured organoid-forming or colony-forming efficiency as a function of donor age or menopausal status in human breast, so whether residual postmenopausal lobules are more, less, or equally clonogenic than premenopausal tissue is simply unknown. The quiescence-depth vocabulary used above, deep versus shallow quiescence with defined molecular markers, derives from mouse lineage-tracing work (Fu, Rios et al. 2017) and has no established human breast counterpart; we retain it as a conceptual scale rather than as a set of measurable human states. And the quiescent human breast stem cell zone was localized to ducts rather than lobules (Villadsen, Fridriksdottir et al. 2007), so siting a reserve compartment in residual TDLUs cannot be justified by that work and requires independent support.

We therefore make a narrower claim than the comparator systems might invite. What the human data support is that persistent postmenopausal lobules retain a proliferation-competent epithelial state embedded in a distinctive immune-stromal architecture (Lue, Stallings-Mann et al. 2026). Whether the epithelium within that state retains genuine progenitor competence, in the sense established for muscle satellite cells or hematopoietic stem cells, is an open question, and an organoid or clonogenicity study stratified by involution status and menopausal timing is the experiment that would settle it. Until that is done, 'reserve' should be read as a structural and microenvironmental description of the persistent lobule, not as a demonstrated claim about the stem cell properties of its epithelium.

Having established the reserve program framework through three comparator systems, we now apply it to ARLI directly. The defining question is whether persistent lobular tissue is passively maintained, cells that simply failed to regress, or actively sustained by microenvironmental signaling in the absence of systemic hormonal input. Two converging lines of evidence bear on this question, and we take them to favor active maintenance without regarding the matter as settled.

***Epidemiologic anchors.*** The natural history of ARLI is characterized by individual variability, biological determination, and clinical consequence independent of standard risk covariates. We note at the outset that variability is not by itself evidence of active maintenance, a point developed below; it establishes only that the phenotype is multifactorially determined. Quantitative TDLU metrics establish that lobular regression is a lifelong process already underway during reproductive years, well before the menopausal transition (Hutson, Cowen et al. 1985, Figueroa, Pfeiffer et al. 2014, Radisky, Visscher et al. 2016). Regression then continues variably for decades, with heterogeneity unexplained by menopausal timing, parity, BMI, or family history (Radisky, Visscher et al. 2016). This constitutes a field-level phenotype: within-woman quadrant concordance in involution scoring is high, supporting ARLI as a whole-breast biological state rather than a focal observation (Vierkant, Hartmann et al. 2009). Critically, involution stasis, the failure to progress across sequential biopsies, carries elevated breast cancer risk independent of absolute involution degree at any single time point, demonstrating that biological trajectory rather than elapsed time governs outcome (Radisky, Visscher et al. 2016).

The magnitude of the risk association is substantial. Postmenopausal women with no involution face approximately threefold elevated risk compared women with complete involution, rising to fivefold with background atypical hyperplasia (Milanese, Hartmann et al. 2006). These effects are independent of mammographic density: the joint phenotype of dense breasts combined with absent involution confers an incidence rate approximately four times that of nondense tissue with complete involution (Ghosh, Vachon et al. 2010). Standardized quantitative TDLU measures replicate this risk gradient with an independent measurement approach: higher TDLU density, span, and acini per TDLU each predict subsequent cancer, with effect estimates ranging from approximately twofold to threefold across extreme quartile contrasts (Figueroa, Pfeiffer et al. 2016). Involution status also intersects with germline risk: a 313-variant polygenic risk score for breast cancer is positively associated with higher TDLU counts, positioning ARLI as a plausible intermediate phenotype linking inherited susceptibility to tissue architecture (Bodelon, Oh et al. 2020). These associations already have clinical applicability: involution extent is incorporated into the BBD-BC absolute risk prediction model with substantial point assignments for absent and partial involution, enabling ARLI-informed risk stratification in women with benign biopsies (Pankratz, Degnim et al. 2015).

The epidemiologic literature is mixed, and this bears on the model directly rather than sitting beside it as a caveat. In a nested case-control study within the Nurses' Health Studies, women whose benign biopsies showed a more involuted lobule-type pattern had lower subsequent breast cancer risk, although the association was attenuated after adjustment for benign histologic category (Baer, Collins et al. 2009). A later large nested case-control analysis in the same parent cohorts, using automated whole-slide quantitative TDLU measures, found no material association between involution metrics and subsequent breast cancer risk (Kensler, Liu et al. 2020). A recent analysis of benign breast disease biopsies likewise found no association between TDLU measures and invasive breast cancer, while senescence-associated morphologic scores derived from the same specimens did predict risk, more strongly among postmenopausal women (Heckenbach, Peila et al. 2025).

Rather than treat this as a limitation to be acknowledged and set aside, we take it as a constraint the model must meet, and the reserve niche framework supplies a specific reason for the inconsistency. If the biological meaning of a senescence-linked tissue state depends on where a woman sits relative to menopause, then cohorts that pool women across the menopausal transition will average over opposing associations and attenuate or abolish the signal. This is not a hypothetical concern. In a longitudinal cohort of 81 women with paired benign breast biopsies, baseline age 45 to 55 with two to ten years of follow-up, senescence-linked inflammatory and innate immune programs reversed the direction of their association with involution pace across the transition, associating with slower remodeling in women traversing menopause and with faster remodeling in women already postmenopausal (Lue, Stallings-Mann et al. 2026). Because the tissue programs were measured before the remodeling they predict, this establishes temporal precedence rather than mere co-occurrence; it does not establish causation, and unmeasured factors could plausibly influence both. The spatial observations we draw on below come from a separate and smaller cross-sectional comparison of postmenopausal noninvoluted and completely involuted tissue within the same cohort, and from imaging subsets of fourteen to sixteen women, which we note wherever those findings are used. Under this model the mixed epidemiologic record is expected rather than anomalous, and it generates a testable prediction: analyses stratified by time from the final menstrual period should recover associations that pooled analyses obscure. It also implies that involution status measured without reference to menopausal timing or microenvironmental state is an incomplete risk variable, which may be the more consequential point for how ARLI is used in risk models.

***The biological activity of persistent lobules.*** A passive target-tissue model accounts for these patterns less readily than an active one, though it is not excluded by them. Postmenopausal breast tissue shows two lobular fates within the same adipose field: a collapsed hypocellular remnant representing complete ARLI alongside an architecturally intact, cellularly active TDLU representing incomplete ARLI (**Fig. 5A**). The most direct functional evidence comes from inflammatory protein profiling of histologically normal breast tissue obtained from pre- and postmenopausal breast cancer patients: elevated levels of IL-6, TNF-α, CRP, COX-2, leptin, SAA1, IL-8, and IL-10 were all inversely associated with complete lobular involution, persisting after age adjustment (Hanna, Dumas et al. 2017) (**Fig. 5B**). That study measured inflammatory protein expression and did not assess senescence by any method; it shows that less involuted lobular tissue is accompanied by a distinct inflammatory and immune-regulatory milieu rather than a neutral excess of residual epithelium, and it does not establish a senescent source. Complementary histologic work likewise argues against passivity. In normal breast tissue, immune cells are predominantly localized to lobules rather than interlobular stroma, with CD8+ cells and dendritic cells closely apposed to the basal epithelial layer; lobules with lobulitis contain significantly higher densities of CD4+, CD8+, CD20+, and CD45+ cells, and in this series, lobulitis was present in 70% of noninvoluted lobules, 43% of partially involuted lobules, and only 3% of completely involuted lobules (Degnim, Brahmbhatt et al. 2014). Single-cell transcriptomic and epigenomic profiling of the aging murine mammary gland provides the broader cellular correlate: in mice, stromal fibroblasts shift toward senescence-associated and cancer-associated states, M2-like macrophages expand as cytotoxic competence declines, and spatial transcriptomics confirms co-localization of aged immune and epithelial cells in situ, with signatures detectable in human breast tumor transcriptomes (Angarola, Sharma et al. 2025). The aged murine mammary microenvironment is not depleted; it is dynamically reorganized within precisely the cellular constituencies, including fibroblasts, macrophages, and immune cells, that the reserve niche model identifies as candidate maintenance signals.

Human data now provide the corresponding observations directly, at a scale that did not exist when this model was first formulated. A single-cell spatial atlas of 527 human breast specimens spanning ages 15 to 86 documents a nonlinear decline in lobule density around menopause together with immune remodeling, including increased M2-like macrophages and granzyme B-positive T cells and reduced B cell and CD8-positive T cell densities in older tissue (Gupta, Lee et al. 2026). Multiplex imaging across 122 women aged 24 to 74 localizes M2-like macrophages specifically to intralobular stroma while peri-epithelial T cell densities decline with age (Zirbes, Joseph et al. 2021). Most directly, noninvoluted postmenopausal lobules carry higher CD45-positive leukocyte densities than completely involuted lobules from the same cohort, alongside higher p16INK4a-positive, Ki67-positive, and TOP2A-positive epithelial densities (Lue, Stallings-Mann et al. 2026).

It should be noted that the atlas-scale data describe a general decline in cellularity and lobule number across the aging breast, which is not in itself evidence of active maintenance and could be read as favoring the opposite. The reserve niche claim is not about the population average; it concerns the lobules that persist, and it stands or falls on whether those lobules differ from involuted ones in organization rather than only in quantity.

Subtype associations provide a second line of support that poorly involuted tissue is biologically distinctive rather than simply more abundant. Among breast cancer cases, reduced involution in adjacent or surrounding normal tissue is preferentially associated with triple-negative and core basal phenotype tumors compared with luminal A cancers, a pattern reported in Polish, Chinese, and Black study populations (Yang, Figueroa et al. 2012, Guo, Sung et al. 2017, Davis Lynn, Lord et al. 2022). This subtype specificity is consistent with a biologically distinct niche state in which inflammatory remodeling, immune regulation, and stromal signaling differ from the microenvironment surrounding fully involuted lobules.

***The postpartum involution analog.*** Postpartum mammary involution provides mechanistic proof-of-concept that the senescence-immune axis is operative and functionally consequential during mammary tissue regression, principally in mice. By contrast with ARLI, postpartum involution is rapid, hormonally triggered, and largely reversible, but the cellular machinery engaged appears homologous, making it the most direct available analog for the niche biology of age-related involution. In mouse models, postpartum involution has been established as a tumor-promotional microenvironment in which collagen remodeling and COX-2 activity are not incidental correlates but functional mediators of progression, framing involution itself as a biologically consequential tissue state rather than a neutral return to baseline (Lyons, O'Brien et al. 2011). Macrophages are required executors of the epithelial clearance program in mice: conditional CSF1R+ macrophage depletion prior to weaning impairs epithelial cell death and alveolar regression, and reconstitution with wild-type or M2-differentiated macrophages fully rescues involution (O'Brien, Martinson et al. 2012). A transiently immunosuppressive milieu accumulates during mid-involution; neutralizing IL-10 reduces tumor growth in postpartum breast cancer models in mice, and comparable immune infiltrates are documented in post-lactational human breast tissue, establishing their relevance beyond rodent models (Martinson, Jindal et al. 2015, Jindal, Narasimhan et al. 2020). The postpartum analog therefore supports the central premise of the ARLI hypothesis: mammary involution can proceed through an immune-stromal program that is transient and resolving in youth, yet biologically potent enough to shape later cancer-relevant tissue states.

The molecular overlap between postlactational and age-related involution is further supported by the observation that plasminogen and phospho-STAT3, biomarkers of active postlactational remodeling, are independently associated with progressive ARLI in serial human biopsy tissue, establishing shared signaling nodes between the two processes (Stallings-Mann,

Heinzen et al. 2017). Most directly, recent work in mice shows that p16$^{INK4a}$-dependent senescence is induced in alveolar luminal cells during the irreversible phase of postpartum involution and contributes to normal tissue remodeling; in a mouse postpartum cancer model, eliminating these involution-associated senescent cells extended tumor latency, indicating that the same senescent program that supports physiologic remodeling can be co-opted when oncogenic events coincide with involution (Chiche, Djoual et al. 2026). If this axis operates during postpartum remodeling in young tissue, it is a natural candidate mechanism for ARLI, with the biology of aging providing candidate reasons why the resolution that normally follows postpartum involution may instead fail.

The limits of this analogy should be stated explicitly, because they are substantial and they constrain what can be carried across. Postpartum involution resolves over days in the mouse and weeks in women, whereas ARLI unfolds over decades; the two processes differ by roughly three orders of magnitude in timescale, and no long-term model currently bridges them. Postpartum involution is also globally synchronous: all alveoli enter regression together under a common systemic trigger, placing every acinus under uniform experimental conditions. Human ARLI is focally heterogeneous, with adjacent lobules in the same microenvironment diverging in the degree to which they regress. A synchronous model can establish which molecular programs are sufficient to execute involution, but it cannot by construction explain why some lobules resist regression while their neighbors do not, and we do not offer it as an explanation for that. We use postpartum involution as a source of candidate mechanisms and testable predictions, not as evidence that those mechanisms operate in human ARLI.

The question of focal divergence has to be addressed with data that resolve individual lobules. Where such data exist, within-breast mosaicism is itself associated with remodeling outcome: the between-lobule variability of epithelial EREG-positive and pSTAT3-positive fields, rather than their mean level, tracks involution pace, consistent with a minority of lobules becoming trapped in persistence-prone configurations rather than a uniform shift across the breast (Lue, Stallings-Mann et al. 2026).

***Menopause as a control point: microenvironmental remodeling and the niche lock.*** Menopause does not simply remove hormonal stimulation from breast tissue; it actively remodels the immune and stromal landscape. The conventional framing, that estrogen withdrawal drives lobular regression and incomplete regression is its consequence, sits awkwardly with the natural history of ARLI. Involution is detectable before the final menstrual period, accelerates around the menopausal transition, continues variably for decades afterward, and shows individual heterogeneity that menopausal timing alone does not explain (Milanese, Hartmann et al. 2006, Radisky, Visscher et al. 2016). We do not treat that unexplained heterogeneity as evidence against a passive account, for reasons set out below; we note only that menopausal timing is an incomplete predictor of outcome. Menopause is better understood as a control point: a transition that reorganizes immune cell composition, remodels macrophage polarization, and reorganizes endocrine inputs to breast stromal biology in ways that may determine whether the programs capable of completing involution are executed.

The immune remodeling that accompanies menopause is more complex than a simple loss-of-immunity model. Postmenopause is associated with increased pro-inflammatory tone and reduced cytotoxic immune function, with contributions from estrogen deprivation in addition to chronological aging (Gameiro, Romao et al. 2010). More recent primary human data sharpen and complicate that picture: postmenopausal women show higher circulating IL-6 and TNF-α, higher total lymphocyte and monocyte counts, and increased exhausted, senescent, and memory T-cell subsets rather than a uniform decline across lymphocyte compartments (Abildgaard, Tingstedt et al. 2020). Postmenopause is accompanied by increased systemic inflammation and altered immune-cell composition and function, with NK activity and lymphocyte-abundance findings varying across studies (Gameiro, Romao et al. 2010, Abildgaard, Tingstedt et al. 2020). Immune competence may therefore change not by uniform decline but by a shift in composition, in which inflammatory tone rises while the balance of effector, memory, and exhausted or senescent states moves, at the same time that senescence-associated burden in breast tissue is increasing (**Fig. 6, locking trajectory**). Whether that compositional shift translates into failed clearance depends, as the human spatial data below indicate, on where effectors are positioned rather than on how many of them remain. This is compounded by the postmenopausal shift from estradiol to estrone: estrone stimulates NFκB-mediated upregulation of CCL2, IL-6, and IL-8 in mammary adipocytes while estradiol opposes this induction, shifting the postmenopausal breast toward a locally pro-inflammatory state that reinforces rather than counteracts SASP-driven niche signaling (Qureshi, Picon-Ruiz et al. 2020). Secondary support for broader menopause-associated immune-stromal rewiring comes from a recent bioRxiv preprint using ovariectomized-mouse single-cell datasets, which inferred estrogen-responsive macrophages as major signaling hubs across several tissues (Iijima, Yamashita et al. 2025); as a preprint reporting inference from mouse data, it is cited as a pointer to a hypothesis rather than as support for the human breast.

Human tissue data indicate that the operative variable here is spatial rather than quantitative, and this changes the claim the model should be making. In a cross-sectional comparison of postmenopausal noninvoluted and completely involuted benign tissue, noninvoluted lobules were not immune-poor. They carried higher CD45-positive leukocyte densities than completely involuted lobules, yet those leukocytes accumulated at the epithelial-stromal boundary and were depleted from the periacinar neighborhoods immediately surrounding p16INK4a-positive and proliferative epithelial foci (Lue, Stallings-Mann et al. 2026). Surveillance is present; positioning consistent with engagement is not. We put it this way advisedly: these measures describe where effectors sit relative to their putative targets and do not assess apoptosis, efferocytosis, or effector function in situ, so what is observed is a configuration compatible with failed clearance rather than a demonstrated failure of it. We therefore frame the failure as execution-limited rather than capacity-limited, which is both more specific than a claim about declining immune competence and more readily falsified, since it predicts a spatial configuration that can be measured directly and can fail to be found.

Menopause therefore defines a critical window during which competing signals for clearance and persistence are both active and a self-sustaining configuration has not yet been established, which is the period of greatest biological plasticity and highest intervention opportunity (**Fig. 6**, **perimenopausal window**). The WHI trials establish timing-dependent biological responsiveness as a general principle of postmenopausal endocrine biology: hormone therapy initiated within versus beyond ten years of the last menses produces fundamentally different risk-benefit profiles, demonstrating that target tissue state at the moment of intervention determines outcome (Manson, Chlebowski et al. 2013). The analogous prediction for the reserve niche is that restoring effector engagement will be most effective before the persistent configuration becomes self-sustaining and hormone-independent; afterward, strategies must target that configuration directly.

A clarification is required at this point, because heterogeneity alone does not decide the question and we do not wish to be read as claiming that it does. That ARLI trajectories vary among women of comparable age and menopausal timing shows only that the phenotype is multifactorially determined. It is equally compatible with passive local variation in hormone exposure, epithelial content, or stromal composition, and the observation that menopausal timing fails to explain the variance does not by itself count against a passive model. Passive persistence remains a legitimate competing hypothesis for the epidemiologic observations considered on their own, and we treat it as one throughout.

What distinguishes the two models is not variance but architecture. A passive model predicts that persistent lobules differ from involuted ones principally in quantity: more residual epithelium, in a microenvironment that is otherwise unremarkable and scales with the tissue present. An active maintenance model predicts a specific and reproducible spatial organization that is not a scaled-up version of normal tissue. The available human data favor the latter. Noninvoluted lobules show tighter intraepithelial clustering of p16INK4a-positive cells at periacinar radii, immune accumulation at the lobular boundary, and reduced leukocyte enrichment in the periacinar neighborhoods around proliferative foci, a combination that a passive excess of residual tissue does not predict (Lue, Stallings-Mann et al. 2026); we note that this comparison rests on a small cross-sectional series of twelve postmenopausal women. We regard this as favoring active maintenance rather than settling the question, and we set out below the observations that would distinguish the two more decisively.

### The Senescent-Immune Reserve Niche Hypothesis

***Cellular senescence in the aging breast: what is established and what is not.*** Claims about senescence require care, because no single marker establishes the senescent state in tissue. Current consensus recommendations call for combined markers and for explicit acknowledgment that senescence is context- and tissue-dependent (Gorgoulis, Adams et al. 2019, Suryadevara, Hudgins et al. 2024). We therefore distinguish throughout between p16INK4a-positive (senescence-associated) cells, which can be identified in situ, and senescent cells defined by multi-marker criteria, which have not been comprehensively characterized in the human breast.

Three lines of evidence bear on senescence in the aging breast, and they differ substantially in strength. First, and most directly, multiplex immunofluorescence in human benign breast tissue shows that noninvoluted postmenopausal lobules carry higher densities of p16INK4a-positive epithelium than completely involuted lobules from the same cohort, together with higher Ki67-positive and TOP2A-positive densities and larger epithelial area (Lue, Stallings-Mann et al. 2026). This is a cross-sectional comparison in twelve postmenopausal women, eight noninvoluted and four completely involuted, and senescence was indexed by epithelial p16 immunopositivity alone rather than by multi-marker criteria. This is the observation on which the present model most directly depends, and it is measured in human tissue stratified by involution status rather than inferred from another species or another organ.

Second, p16INK4a induction is an established feature of human mammary epithelium. p16 accumulates at the first senescence arrest in cultured human mammary epithelial cells, and its inactivation extends growth capacity (Brenner, Stampfer et al. 1998); p16INK4a promoter methylation occurs in vivo in histologically normal human mammary epithelium (Holst, Nuovo et al. 2003); and in situ p16 immunohistochemistry in archival human breast biopsies shows p16 expression increasing with age at biopsy (Barr Fritcher, Degnim et al. 2011). Deep-learning assessment of senescence-associated nuclear morphology, applied to normal breast tissue from healthy volunteer donors and to benign breast disease biopsies, associates senescence burden with subsequent breast cancer risk, more strongly in postmenopausal women (Heckenbach, Powell et al. 2024, Heckenbach, Peila et al. 2025). These are morphologic surrogates rather than marker-defined senescence, and their strongest signal was in adipose rather than epithelial compartments; both qualifications matter for how much weight the finding can carry.

Third, single-cell transcriptomic and epigenomic profiling of aged mammary gland shows stromal fibroblasts upregulating Cdkn2a and Cdkn1a and shifting toward senescence-associated and cancer-associated states, with expansion of M2-like macrophages and of T cell subsets bearing exhaustion markers (Angarola, Sharma et al. 2025). That study profiled mouse mammary gland at 3 and 18 months of age; no human breast tissue was profiled, and human data entered only as a comparison of transcriptional signatures against existing tumor datasets. It should therefore be read as a description of the aged murine gland, from which predictions for the human breast follow but evidence does not.

What is not established should be stated as plainly as what is. Two recent surveys of senescent-cell markers across human tissues did not include breast (Idda, McClusky et al. 2020, Saul, Jurk et al. 2025), so there is no human atlas of p16 and p21 burden in the breast comparable to those available for skin or lung. No study has measured senescence-associated β-galactosidase activity or applied multi-marker senescence criteria in situ in normal human breast stratified by involution status. The senescent burden of the aging human breast is therefore better described as indicated by p16INK4a-positive epithelial foci and by morphologic and transcriptional surrogates than as comprehensively quantified, and the model that follows is built on that more limited foundation.

The functional output of these cells is the SASP, a context-dependent secretome comprising inflammatory cytokines, matrix metalloproteinases, and, critically, EGFR family ligands including amphiregulin (AREG) and epiregulin (EREG) (Basisty, Kale et al. 2020). Whether SASP is transiently deployed during normal remodeling or chronically sustained in aging tissue determines its functional consequences (Birch and Gil 2020). This temporal duality maps directly onto the distinction between complete and incomplete ARLI: the postpartum involution niche represents transient, resolving senescent signaling; the aged breast microenvironment represents chronic, non-resolving SASP accumulation (Lue and Radisky 2025).

***The hypothesis: formal statement.*** We propose that incomplete ARLI represents the postmenopausal persistence of a reserve tissue program that was actively maintained during reproductive life, sustained now not by ovarian hormones but by local paracrine signaling that substitutes for the systemic cues that have been withdrawn. The core proposition is as follows: persistent senescence-associated paracrine signaling, combined with a spatial reorganization of immune effectors that leaves them recruited to the lobule but not engaged with their targets, maintains residual epithelial units in a survival-primed, incompletely involuted state (**Fig. 7**). We state the proposition in this form deliberately: in human tissue, noninvoluted lobules are immune-rich rather than immune-poor, and the same senescence-linked programs that mark stagnation during the menopausal transition mark accelerated regression after it (Lue, Stallings-Mann et al. 2026). The distinction, then, is one of execution rather than capacity, of whether effectors are positioned to engage p16INK4a-positive targets rather than how many effectors are present or how much inflammation the tissue carries. We retain the term niche lock for the resulting state, defining it here as the spatial uncoupling of surveillance from its targets rather than as a loss of immune capacity. Under this model, incomplete ARLI is not a failure to involute but a maintenance program running without its resolution signal, in which surveillance is recruited but clearance is not executed. Three mechanistic arms are proposed to generate this state. Each draws on published evidence, and we indicate for each what is now supported in human breast tissue and what remains a proposal.

***The reserve state, defined operationally.*** Because the assessment of this work identified the absence of operational definitions as its principal weakness, we state here what would count as the proposed state in human breast tissue, in terms that can be measured and that can fail to be satisfied. A persistent lobule qualifies when three conditions hold together. First, the epithelium is proliferation-competent, indexed by Ki67-positive epithelial nuclear density exceeding that of completely involuted lobules. Second, senescence-associated epithelium is enriched, indexed by p16INK4a-positive epithelial density and by tighter intraepithelial clustering of p16INK4a-positive cells at periacinar radii. Third, and definitionally, immune effectors are present but not engaged: leukocyte density equals or exceeds that of involuted lobules, while leukocytes accumulate within approximately 20 μm of the epithelial-stromal interface and are depleted from the periacinar neighborhoods surrounding p16INK4a-positive and proliferative foci.

TThese criteria are operationalized with explicit thresholds in the cohort study cited above, where interface coupling was scored by whether every boundary leukocyte had a p16INK4a-positive epithelial cell within 20 μm, and individual lobules were classified by quintile cutoffs of effector abundance and effector-to-target distance (Lue, Stallings-Mann et al. 2026). Two limitations of that operationalization should be carried forward with it. The interface-coupling interaction across menopausal trajectory was directionally consistent but not statistically significant in a fourteen-woman imaging subset, so the third criterion is provisionally specified rather than validated. And these metrics quantify the spatial positioning of effectors relative to their putative targets; they do not measure apoptosis, efferocytosis, or effector function in situ, so what they describe is a clearance-associated configuration rather than a demonstrated clearance event.

***Arm 1: SASP as a paracrine maintenance signal for residual epithelium***. AREG is among the most consistently elevated factors across SASP profiles (Basisty, Kale et al. 2020). In the mouse mammary gland, AREG is the primary paracrine mediator of estrogen-driven ductal morphogenesis, functioning through stromal estrogen receptor alpha to drive EGFR-dependent epithelial proliferation (Ciarloni, Mallepell et al. 2007). Its presence as a canonical SASP factor positions senescent stromal cells as a candidate substitute source of epithelial maintenance signal after ovarian hormone withdrawal.

IL-6/STAT3 signaling provides a second paracrine arm with a specifically mammary-relevant paradox. In mice, acute STAT3 activation is required for epithelial apoptosis during postpartum involution, and conditional Stat3 deletion dramatically delays involution and suppresses cell death, with failure of the downstream survival-signal sequestrant IGFBP-5 to accumulate (Chapman, Lourenco et al. 1999, Hughes, Wickenden et al. 2012). Yet the same IL-6/STAT3 axis functions as a chronic survival and stemness pathway in cancer and aberrant epithelial states. Under chronic low-level SASP-derived IL-6, without the acute milk-stasis context of postpartum remodeling, STAT3 signaling in residual lobular epithelium may favor survival rather than clearance, a context-dependence consistent with the well-characterized pleiotropy of this pathway.

Human data place this arm on firmer ground while also complicating it in an instructive way. In human breast lobules, epithelial fields positive for the EGFR ligand EREG show a menopause-conditioned relationship with remodeling: proliferative cells embedded within an EREG-rich field mark stalled remodeling in women traversing menopause, but accelerated regression in women already postmenopausal (Lue, Stallings-Mann et al. 2026). The EGFR ligand arm is therefore active in the human breast and associated with involution trajectory, but its directionality is conditioned by menopausal timing. A paracrine signal that maintains epithelium in one physiologic context and accompanies its clearance in another is not adequately described as a maintenance factor, and we do not describe it as one.

***Arm 2: Convergent mechanisms by which senescence-associated cells evade clearance.*** Four convergent mechanisms are proposed to allow persistent senescence-associated cells to evade immune clearance, all characterized in human tissues or experimental models outside the breast; their operation specifically within the ARLI lobular niche remains to be demonstrated. First, MMP-dependent shedding of NKG2D ligands (MICA, MICB, ULBPs) by senescent cells removes the activating signal for NK and CD8 T cell killing, while accumulating soluble MICA further downregulates NKG2D in a paracrine feed-forward loop (Munoz, Yannone et al. 2019). Second, SASP-derived IL-6 induces HLA-E upregulation on senescent cells, engaging the inhibitory receptor NKG2A on NK cells and terminally differentiated CD8 T cells, an axis reversible by NKG2A blockade (Pereira, Devine et al. 2019). Third, p16-dependent CDK4/6 inhibition stabilizes PD-L1 by blocking its proteasomal degradation; p16-expressing senescent macrophages thereby create an immunosuppressive environment that amplifies senescent burden through a self-reinforcing loop (Majewska, Agrawal et al. 2024). Fourth, senescent cells upregulate CD47, the canonical anti-phagocytic signal, suppressing macrophage efferocytosis in a contact-dependent, SASP-independent manner that extends to failure of bystander apoptotic corpse removal (Schloesser, Lindenthal et al. 2023).

These four mechanisms remain hypotheses for ARLI. None has been demonstrated in the lobular niche, and we do not present them as established components of the model. They are, however, motivated by an observed human phenotype. If immune cells accumulate at the lobular boundary yet fail to engage p16INK4a-positive foci within the same lobule (Lue, Stallings-Mann et al. 2026), then mechanisms permitting senescence-associated cells to persist despite immune-cell presence become the relevant candidate class, and the four above are its best-characterized members. They are also not the only candidates: a purely architectural explanation, in which extracellular matrix barriers or deficient chemokine gradients prevent effectors from reaching targets they would otherwise engage, would produce the same spatial observation without any of these molecular mechanisms operating. Distinguishing among them is a matter for spatial profiling of matched tissue rather than for further inference.

**Arm *3: The macrophage as integrating niche hub.*** The immune evasion mechanisms of Arm 2 converge on macrophages as the primary cellular integrator. Tissue-resident macrophages are the principal executors of epithelial clearance during postpartum involution in mice (O'Brien, Martinson et al. 2012), and a recent bioRxiv preprint inferred cross-tissue signaling networks from ovariectomized-mouse single-cell datasets and identified estrogen-responsive macrophages as major signaling hubs across multiple tissues (Iijima, Yamashita et al. 2025). The liver fibrosis system provides a mechanistic precedent from another organ: NK-mediated clearance of senescent hepatic stellate cells is causally required for fibrosis resolution, and its impairment allows SASP accumulation and progressive fibrotic remodeling (Krizhanovsky, Yon et al. 2008). Under the reserve niche model, the proposed shift is from clearance-executing to niche-sustaining: macrophages remain present in and around lobular structures, and may retain close contact with them, while failing to engage and eliminate senescence-associated cells and instead providing trophic signals that sustain residual epithelium. Stated in this form the proposal concerns positioning and engagement rather than macrophage abundance, and it is falsifiable by measuring both. The immunosuppressive postpartum macrophage phenotype, including IL-10 production, Foxp3+ Treg co-recruitment, and M2-like polarization (Martinson, Jindal et al. 2015), may be aberrantly sustained in the aged postmenopausal breast rather than resolving as it normally does.

Direct human evidence now bears on this arm. In human breast tissue, the density of M2-like CD68-positive CD206-positive macrophages and the degree to which their distribution tracks p16INK4a-positive epithelial foci across lobules both show menopause-conditioned associations with involution pace, with macrophage targeting of senescence-associated epithelium accompanying faster involution after menopause and the opposite association during the transition (Lue, Stallings-Mann et al. 2026). Macrophage organization relative to senescence-associated targets is therefore associated with tissue outcome in the human breast. Whether that organization is causal, and whether repolarization would change involution trajectory, remains untested.

***Experimental predictions and the logic of testing.*** The three arms generate orthogonal, testable predictions. Arm 1 predicts that senescent cells should be spatially associated with persistent lobular epithelium, and that senolytic clearance should reduce epithelial survival signaling and shift lobular persistence. Arm 2 predicts that senescent cells adjacent to persistent lobules should display immune evasion program upregulation, including HLA-E, PD-L1, CD47, and that targeted clearance restoration should alter lobular trajectory. Arm 3 predicts that macrophage efferocytosis competence and polarization should differ systematically between women with complete versus incomplete ARLI, and that macrophage depletion or repolarization in aged murine mammary tissue should alter residual lobular architecture. These predictions are testable with existing tools: p16-reporter mouse models for senescence visualization and inducible clearance, spatial transcriptomics for lobule-senescent cell co-localization, organoid co-culture for paracrine signal dissection, and macrophage depletion or adoptive transfer for niche hub interrogation. The primary experimental gap, and the first-priority test, is spatial association of senescent cells with persistent lobular structures in human tissue correlated with involution status. Its outcome will determine whether the mechanistic arms warrant the more resource-intensive functional models.

Two of these predictions have since been partially tested, and the outcome is reported here rather than left as a forward-looking claim. The first-priority test identified above, spatial association of senescence-associated cells with persistent lobular structures in human tissue correlated with involution status, has been performed: p16INK4a-positive epithelial density is higher in noninvoluted than in completely involuted postmenopausal lobules, and immune cells within those lobules accumulate at the epithelial-stromal boundary while being depleted from periacinar neighborhoods (Lue, Stallings-Mann et al. 2026). The macrophage prediction of Arm 3 has likewise received support, in a fourteen-woman imaging subset analyzed by exact permutation testing. What remains untested is the causal core of all three arms: whether removing senescence-associated cells, restoring effector engagement, or repolarizing macrophages alters involution trajectory. Each requires intervention in a long-term model, and the acute postpartum models currently available cannot substitute for one.

**Translational Implications**

Three conditions must be met before any of this becomes actionable. Senescence-associated cells must be shown to be enriched near persistent lobules; the relevant cell types and immune states must be identified; and those cells must be shown to influence involution trajectory or cancer risk causally. The first condition is now supported in human tissue, the second is partially addressed, and the third is not met. What follows should be read as a map of what would become testable if the third condition were satisfied, not as a proposal for near-term intervention, and none of the agents named below should be understood as a candidate for clinical use in this setting on present evidence.

***Risk stratification: from morphometry to microenvironmental biomarkers.*** ARLI-based risk assessment currently rests on morphometric parameters, primarily TDLU count, lobular density, and degree of involution; these describe tissue architecture but are silent about the mechanisms maintaining it. If incomplete ARLI represents an actively sustained senescent-immune niche state, the informative signal is not the count of persistent lobules but the molecular and spatial configuration of the lobular microenvironment: senescence-associated burden, SASP composition, and, critically, whether immune effectors are positioned to engage their targets. Candidate tissue biomarkers follow from the hypothesis. $p16^{INK4a}$ burden in epithelial and stromal compartments provides a spatially resolved distribution of senescence-associated cells; the approach was validated as an aging biomarker across rodent tissues, which did not include mammary gland (Krishnamurthy, Torrice et al. 2004), and has since been applied to human breast lobules stratified by involution status (Lue, Stallings-Mann et al. 2026). SASP cytokine profiles, particularly IL-6, IL-8, AREG, and EREG, report on paracrine epithelial signaling. Immune markers including NKG2A and HLA-E expression, PD-L1 abundance on senescence-associated populations, and macrophage efferocytosis capacity report on the convergent evasion mechanisms proposed in Arm 2 (Munoz, Yannone et al. 2019, Pereira, Devine et al. 2019, Schloesser, Lindenthal et al. 2023, Majewska, Agrawal et al. 2024). Under the execution-limited formulation, spatial metrics may be more informative than abundance metrics: measures of whether leukocytes at the lobular boundary are uniformly adjacent to p16INK4a-positive epithelium, and of how tightly senescence-associated foci cluster within the lobule, separate persistent from involuted tissue in human breast where abundance alone does not (Lue, Stallings-Mann et al. 2026). We note that the corresponding longitudinal interaction, whether interface coupling predicts subsequent involution pace differently across the menopausal transition, was directionally consistent but did not reach significance in a fourteen-woman imaging subset, so these measures are candidate biomarkers rather than validated ones. Non-invasive blood-based proxies are also motivated: GDF15 and SERPINE1 show the strongest associations with biological age in the SASP proteome (Basisty, Kale et al. 2020), offering longitudinal monitoring in low-burden paradigms. These biomarkers are likely most informative during the perimenopausal window, arguing for risk assessment timed to the transition rather than to fixed postmenopausal intervals.

***Therapeutic implications and the primacy of timing.*** Three mechanistically distinct intervention arms map onto the locking trajectory, and their plausibility is likely to be strongly timing-dependent (**Fig. 8**). Senolytics provide the most direct conceptual route to disrupting an established niche lock (**Fig. 8, Arm 3**). Human pilot studies in idiopathic pulmonary fibrosis have established feasibility and tolerability for dasatinib plus quercetin, offering translational precedent for targeting senescent burden in a chronic fibrotic microenvironment (Justice, Nambiar et al. 2019, Nambiar, Kellogg et al. 2023). Navitoclax is an additional senolytic candidate with demonstrated activity against several senescent cell types, although its effects are cell-type restricted and its relevance to ARLI is entirely hypothetical (Zhu, Tchkonia et al. 2016).

A critical caveat applies here, and it deserves more weight than a caveat customarily receives. Transient p16INK4a-dependent senescence is required for normal postpartum mammary involution: in mice, alveolar luminal cell-specific p16 deletion and senolytic treatment with ABT-263 each delayed alveolar regression and adipocyte repopulation (Chiche, Djoual et al. 2026). In the same experiments, ABT-263 also reduced the incidence of palpable tumors, delayed tumor onset, and suppressed lung metastasis. The remodeling and tumor-promoting consequences of senescent cells therefore dissociated under a single intervention. We read this as informative rather than contradictory: it indicates that senescent cells in involuting mammary tissue serve a remodeling function and a tumor-permissive function that are separable, and that an intervention can impair one while relieving the other.

It also means that senolytic timing is not a secondary consideration but the primary one. Poorly timed senolytic therapy risks disrupting physiologic remodeling in premenopausal women or during the peri-transition period. Equally important, these are acute rodent experiments conducted over days, and they cannot be directly extrapolated to a chronic process unfolding over decades in women; the long-term models that would justify such an extrapolation do not exist. Senolytics are therefore best viewed as hypothetical candidates for an already established persistent configuration, not as preventive agents, and we make no claim that the postpartum result supports their use in ARLI.

Immune restoration strategies offer complementary approaches tied to specific clearance mechanisms (**Fig. 8, Arm 2**). NKG2A blockade reverses HLA-E-mediated NK and CD8 T cell inhibition of senescent cells (Pereira, Devine et al. 2019); MerTK agonism or CD47/SIRPα targeting addresses the innate efferocytosis arm (Schloesser, Lindenthal et al. 2023); NK adoptive transfer could restore cytotoxic surveillance without systemic senolytic off-target risk. SASP attenuation through senomorphics, EGFR pathway modulation, or IL-6/STAT3 axis targeting offers a third strategy: reducing paracrine survival signals to residual epithelium without necessarily eliminating senescent cells (**Fig. 8, Arm 1**).

The central translational principle is that context-dependence is not a complication of trial design: it is the signal (**Fig. 8, footer**). Perimenopausal women are the appropriate population for interventions aimed at preventing niche lock formation; postmenopausal women with established incomplete ARLI are the appropriate population for interventions that seek to break an already self-sustaining lock. Treating menopausal timing as a covariate to adjust for, rather than a primary stratification variable, risks obscuring the very interaction that is most likely to determine efficacy.

***The breast as a model system for transition-biology research.*** ARLI occupies an unusual position among human aging processes: triggered by a defined physiologic event, producing a measurable structural outcome, following a decadal natural history with documented individual variability, and accessible to serial tissue sampling through core biopsy; a combination unavailable for comparable aging questions in liver, lung, bone marrow, or kidney. If perimenopausal immune state can be shown to predict long-term tissue fate in the breast, the principles generated, including critical windows of biological plasticity, context-dependent interpretation of inflammatory signals, timing-dependent responsiveness of tissue fate, are unlikely to be breast-specific, and the findings will carry implications for transition-biology research well beyond breast cancer risk.

**Conclusions and Future Directions**

Three questions define the most consequential experimental frontier. The first and most foundational question is what kind of biological state incomplete ARLI actually represents. Throughout this review we have used "adaptive reserve" in an evolutionary rather than a functional sense: the lobular maintenance program was likely selected during reproductive years, when keeping epithelial capacity available across cycles of partial involution had genuine biological value. The postmenopausal breast is not awaiting reactivation for lactation; the program is not serving a current purpose. What persists postmenopausally is more accurately described as a maintenance program running without its resolution signal, the hormonal and immune cues that would ordinarily terminate it having been progressively withdrawn across the menopausal transition. Whether the epithelium retained within that program preserves any meaningful progenitor properties, or whether it has become purely a senescent-adjacent scaffold sustained by niche signaling without functional reserve competence, is the consequential experimental question. The answer matters for therapeutic strategy: progenitor-competent epithelium and inert inflammatory scaffold present different molecular targets, and the reserve niche framework predicts different intervention points depending on which state predominates.

The second question concerns which cell types are the primary drivers of the senescent niche. The hypothesis invokes SASP-producing stromal cells as the paracrine maintenance signal for adjacent epithelium, but the dominant senescent population, whether mammary epithelial cell, fibroblast, adipose stromal cell, vascular endothelial cell, or immune cell, has not been established for the aging human breast. The SASP components most relevant to each mechanistic arm, AREG and EREG for epithelial maintenance, MMPs for NKG2D ligand shedding, IL-6 for HLA-E induction, may originate from distinct cellular sources whose spatial relationship to persistent lobules has yet to be mapped.

The third question concerns immune spatial organization within the persistent lobule. The reserve niche model requires that immune surveillance cells fail to eliminate senescence-associated cells in the lobular microenvironment, and human spatial data are consistent with that failure: effectors are present in greater numbers than in involuted tissue, yet accumulate at the epithelial-stromal boundary and are depleted from the periacinar neighborhoods around p16INK4a-positive foci (Lue, Stallings-Mann et al. 2026). Positioning is not the same as function, and whether these effectors would kill their targets on contact is untested. What remains unresolved is why. Active exclusion through inhibitory ligand expression, extracellular matrix barriers to trafficking, deficient chemokine gradient formation, and failure of effector function after contact all remain live possibilities, and they are distinguishable from one another by direct measurement. The mechanistic answer matters for therapeutic design: strategies targeting cytotoxic effector function, such as NKG2A blockade or NK cell transfer, will fail if the primary barrier is physical access rather than functional suppression. Spatial transcriptomics applied to matched complete and incomplete ARLI tissue with multiplexed immune phenotyping, extended to include efferocytosis and effector functional readouts that current panels do not resolve, is the approach most directly suited to settling this.

A fourth question extends the reserve niche logic to a connected epidemiologic puzzle: the divergent cancer risk profiles of early versus late pregnancy. Early pregnancy confers durable long-term protection against breast cancer risk, while late pregnancy does not (MacMahon, Cole et al. 1970, Ewertz, Duffy et al. 1990, Collaborative Group on Hormonal Factors in Breast 2002, Albrektsen, Heuch et al. 2005, Lue and Radisky 2025). The reserve niche model generates a testable mechanistic explanation for this asymmetry. Late pregnancy occurs as women approach the perimenopausal transition, the window during which senescence-associated burden is rising and, under the model proposed here, the spatial coupling between immune effectors and their targets is least reliable. Postpartum involution in mice requires transient p16$^{INK4a}$-dependent senescence for epithelial clearance and tissue remodeling (Chiche, Djoual et al. 2026); if a woman enters postpartum remodeling against a background of already-elevated senescence-associated burden and impaired effector engagement, this transient senescent signal may fail to resolve normally. The predicted result is a hybrid state in which elements of postpartum senescent signaling merge with the nascent conditions of an age-related niche lock, impairing completion of postpartum involution and leaving behind a microenvironment with niche lock characteristics. Under this model, the attenuated protective effect of late pregnancy is not simply a function of epithelial differentiation status but reflects the immune-senescent context in which postpartum remodeling must occur, a context that conditions whether the transient senescent program resolves or entrains. This prediction is experimentally testable by comparing postpartum involution trajectory and the spatial coupling of innate effectors to p16INK4a-positive epithelium in women delivering at different ages, stratified by perimenopausal senescence-associated burden at the time of delivery.

If the immune evasion mechanisms already present in the reserve niche, PD-L1 stabilization, HLA-E expression, CD47 upregulation, provide early immune escape scaffolding for emerging tumor cells, the framework will connect ARLI epidemiology to specific breast cancer subtype biology in a mechanistically grounded way, opening new lines of thinking about early interception timed to the niche rather than to established premalignant lesions. The reserve niche hypothesis offers a reorientation of how breast cancer risk is understood: from a static count of persistent tissue to a dynamic, targetable microenvironmental program; one whose most consequential biology unfolds in the perimenopausal window, before the lock is set.

**Figure 1. Skeletal muscle satellite cells establish the cellular reserve program logic: quiescence, niche maintenance, and aging-driven senescent conversion**

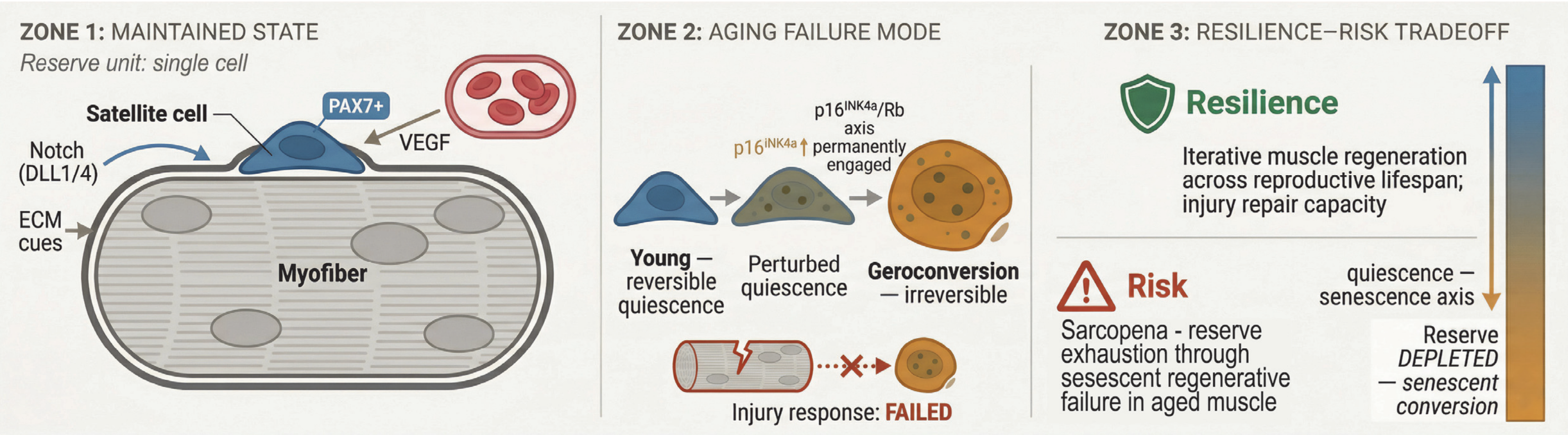

**Figure 1. Skeletal muscle satellite cells establish the cellular reserve program logic: quiescence, niche maintenance, and aging-driven senescent conversion.** PAX7+ satellite cells (steel blue) reside between the myofiber sarcolemma and basal lamina, maintained in reversible G0 quiescence by Notch ligands, extracellular matrix cues, and paracrine VEGF from adjacent vasculature (**Zone 1**). Niche-enforced quiescence preserves reactivation competence across multiple cycles of injury and repair. With aging, progressive engagement of the $p16^{INK4a}$/Rb axis drives an irreversible transition through perturbed quiescence to geroconversion, a cell-intrinsic senescent state from which reactivation is no longer possible and injury-induced regeneration fails (**Zone 2**). The quiescence-to-senescence spectrum (right axis) illustrates this progression: young satellite cells occupy the deep-quiescence end; aging drives movement toward irreversible senescence, depleting the reserve. **Zone 3** encodes the resilience-risk tradeoff: iterative regenerative capacity across the reproductive lifespan, at the cost of sarcopenic reserve exhaustion through senescent conversion in aged muscle. The core molecular logic, $p16^{INK4a}$/Rb-dependent quiescence maintenance, niche-enforced reversibility, and aging-driven senescent conversion, recurs across reserve tissue programs and is directly applicable to the breast ARLI system.

**Figure 2. Hematopoietic stem cells establish the somatic evolutionary horizon principle: cancer risk without pre-malignant conversion in a long-lived maintained reserve**

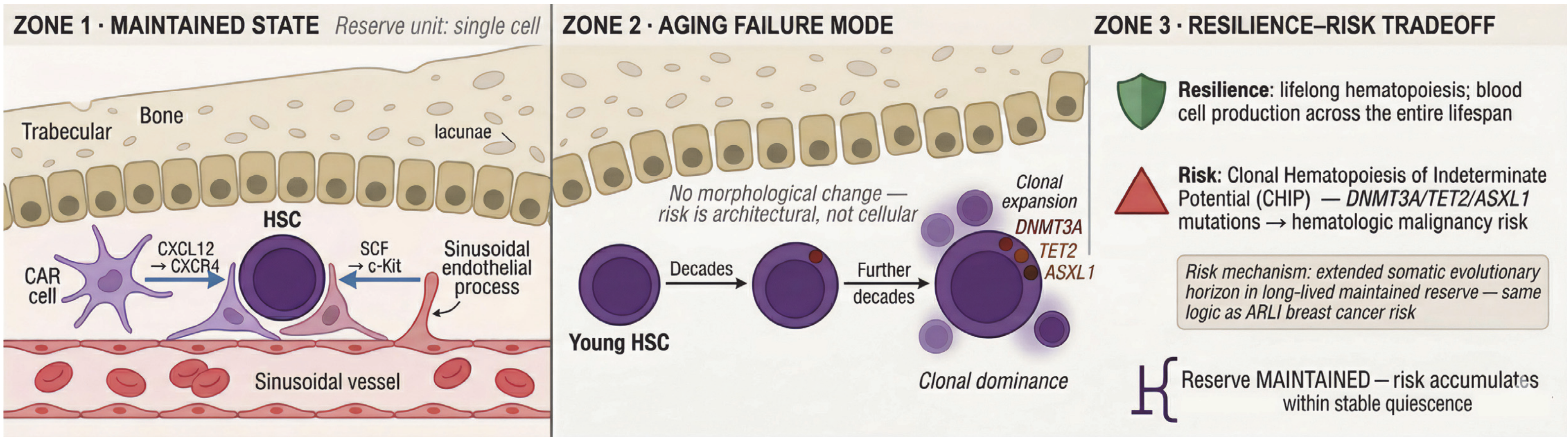

**Figure 2. Hematopoietic stem cells establish the somatic evolutionary horizon principle: cancer risk without pre-malignant conversion in a long-lived maintained reserve.** HSCs (deep violet) are maintained in stable quiescence within the endosteal niche by CXCL12/CXCR4 retention signals from CAR cells and SCF/c-Kit signaling from sinusoidal endothelium (**Zone 1**). Unlike skeletal muscle satellite cells, aging does not drive HSCs toward senescent conversion, the reserve persists in stable quiescence across decades with no morphological change. Risk accumulates instead through somatic evolution within the maintained progenitor pool: clones bearing driver mutations in DNMT3A, TET2, and ASXL1 expand gradually to clonal dominance without producing overt disease (CHIP; **Zone 2**). Clonal dominance is invisible at the cellular level; risk is architectural, not cellular. **Zone 3** encodes the resilience-risk tradeoff: lifelong hematopoiesis across the entire lifespan, at the cost of hematologic malignancy risk through an extended somatic evolutionary horizon operating within the stably maintained reserve. The principle is directly applicable to incomplete ARLI: a maintained, proliferation-competent tissue unit confers cancer risk through the extended opportunity it provides for somatic evolution, independent of pre-malignant conversion of reserve cells themselves.

**Figure 3. The postmenopausal endometrium illustrates passive reserve dormancy and risk through aberrant reactivation**

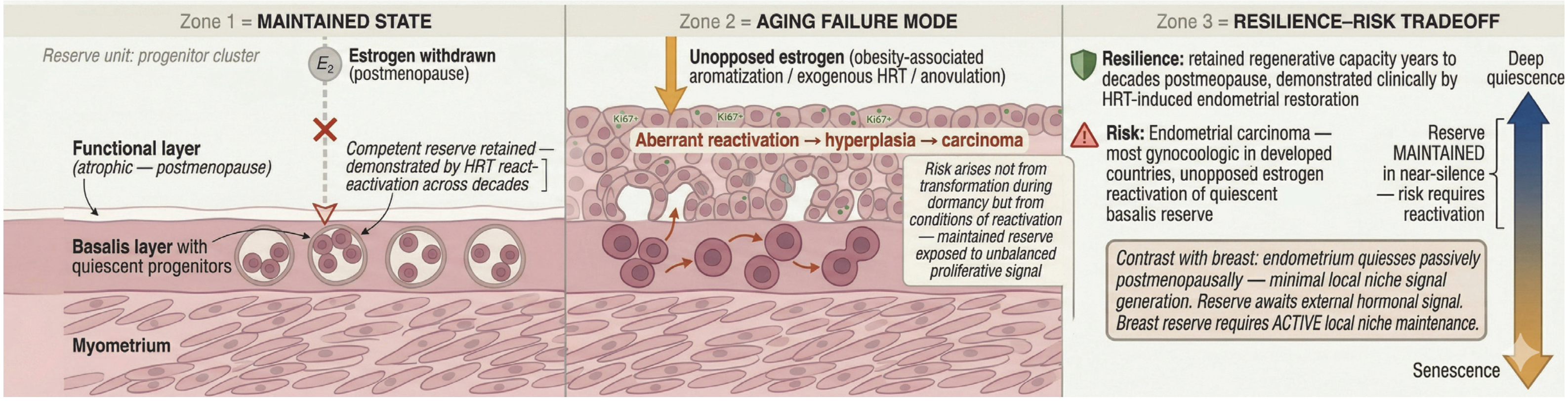

**Figure 3. The postmenopausal endometrium illustrates passive reserve dormancy and risk through aberrant reactivation.** Basalis progenitor cells clustered in gland bases at the myometrium interface persist in deep quiescence following menopause, maintained without hormonal stimulation in atrophic but architecturally intact tissue for years to decades (**Zone 1**). The functional competence of this dormant reserve is demonstrated clinically by the rapid restoration of pre-menopausal endometrial thickness following exogenous estrogen administration, as the quiescent reserve retains full reactivation capacity regardless of time elapsed since menopause. Risk arises not from transformation during dormancy but from aberrant reactivation: unopposed estrogen from obesity-associated aromatization or exogenous hormone regimens triggers abnormal proliferation from the maintained reserve, driving endometrial hyperplasia and carcinoma (**Zone 2**). The quiescence-to-senescence spectrum (right axis) positions endometrial progenitors at the deep-quiescence end with minimal aging-driven movement; the reserve awaits an external hormonal signal rather than generating internal niche dynamics. **Zone 3** encodes the resilience-risk tradeoff: retained regenerative capacity years to decades postmenopause, at the cost of carcinoma risk through unopposed estrogen reactivation of the quiescent basalis reserve. Critically, the postmenopausal endometrium quiesces passively; minimal active local niche signal generation is required, and the reserve awaits an external hormonal cue. This is in direct contrast to the breast reserve, which requires active local paracrine maintenance in the absence of systemic hormonal input and is sustained rather than resolved by its immune-stromal microenvironment.

## Figure 4. The breast TDLU reserve operates at architectural rather than cellular scale, requiring active local niche maintenance in the absence of systemic hormonal input

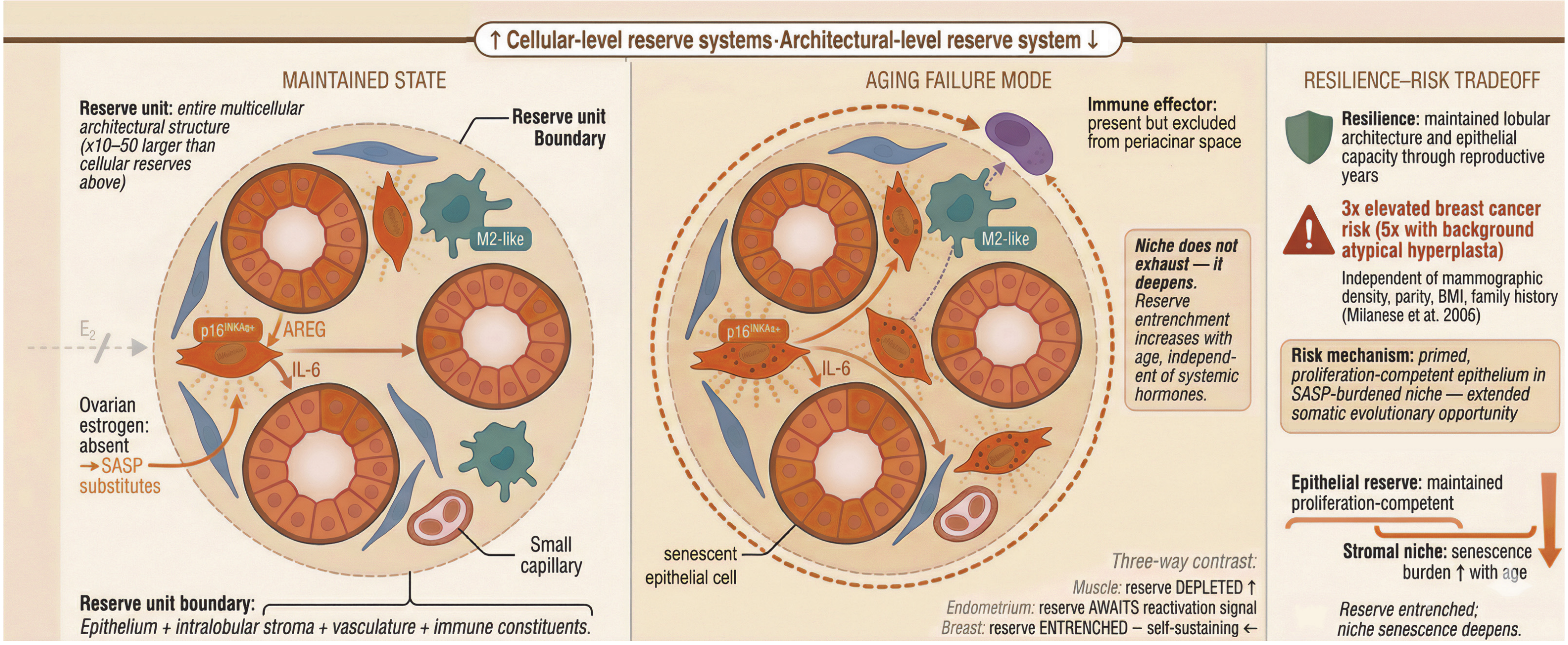

**Figure 4. The breast TDLU reserve operates at architectural rather than cellular scale, requiring active local niche maintenance in the absence of systemic hormonal input.** The unit of reserve in incomplete ARLI is an entire functional terminal duct lobular unit: epithelium, intralobular stroma, vasculature, and immune constituents together (dashed boundary), rather than a single stem-cell compartment. This architectural scale implies a requirement not dependent on canonical cell-level reserves: active local paracrine signal generation is needed to sustain the unit after systemic ovarian input has withdrawn. In the maintained state (left), senescent stromal cells are proposed to substitute in part for withdrawn endocrine support through SASP-derived epithelial maintenance signals such as AREG and IL-6, while macrophages and other immune/stromal components contribute to a trophic microenvironment. The aging failure mode (right) is distinct from the comparator systems: rather than reserve depletion through senescent collapse or passive dormancy awaiting reactivation, the breast niche becomes progressively self-sustaining as senescent burden accumulates and immune clearance fails despite continued immune-cell presence. The resilience-risk tradeoff is maintenance of lobular architecture and epithelial capacity across reproductive years at the cost of a chronically survival-permissive microenvironment that extends the somatic evolutionary horizon of the tissue unit.

# Figure 5. Schematic summary of evidence that incomplete age-related lobular involution is a biologically active maintained tissue state

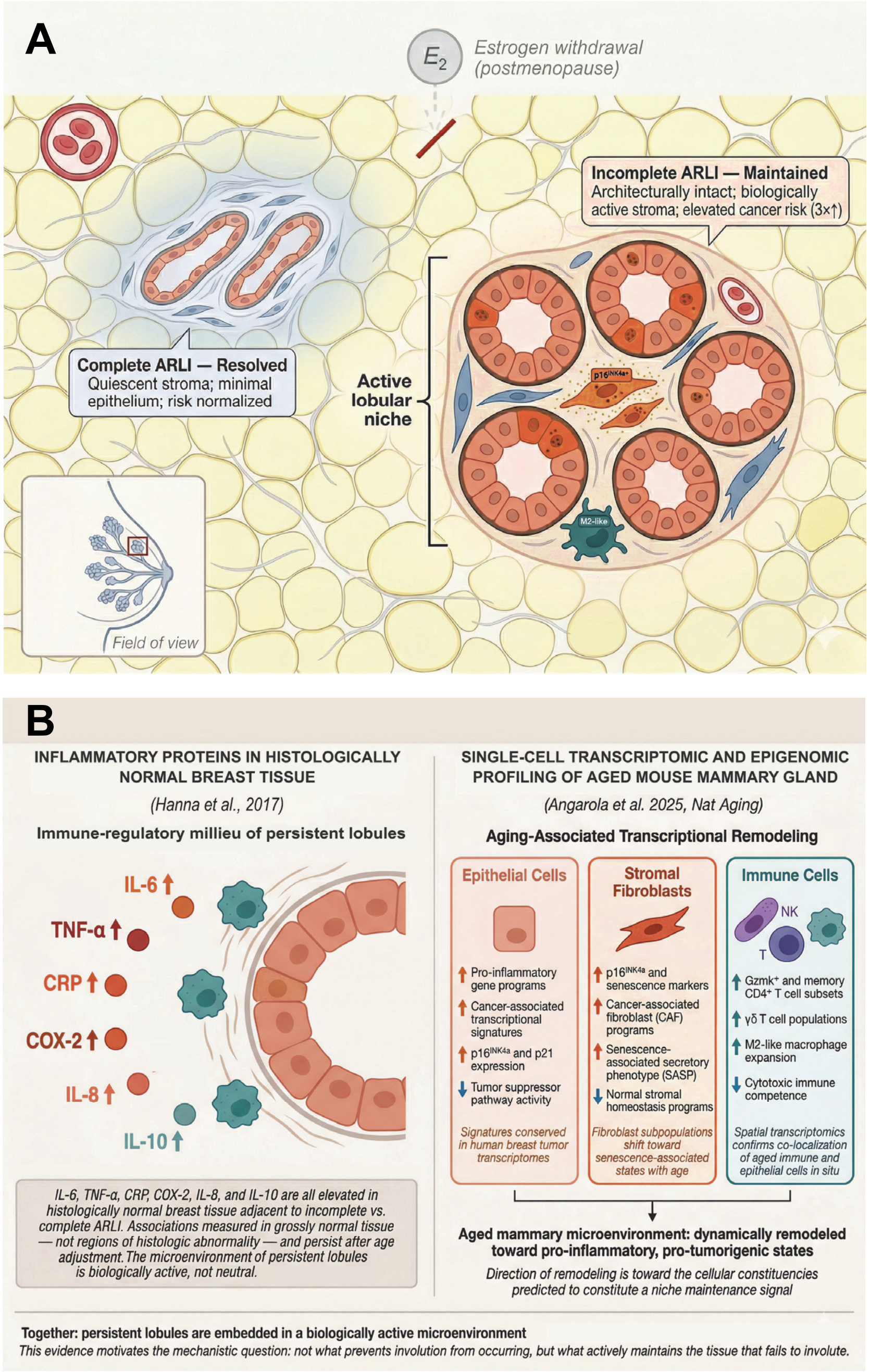

**Figure 5. Schematic summary of evidence that incomplete age-related lobular involution is a biologically active maintained tissue state. (A)** Wide-field cross-section of postmenopausal breast tissue showing two lobular fates within the same adipose field. Complete ARLI (left): a collapsed, hypocellular lobular remnant with quiescent stroma embedded in interlobular fat. Incomplete ARLI (right): an architecturally intact terminal duct lobular unit with organized acinar profiles and a biologically active intralobular stroma containing fibroblasts, macrophages, and persistent epithelial structures. Ovarian estrogen withdrawal is indicated above; the lobular niche persists despite it. Anatomical inset indicates field of view within the mammary ductal tree. **(B)** Two lines of evidence that persistent lobules are embedded in a biologically active microenvironment. Left: in histologically normal human breast tissue from breast cancer patients, pro-inflammatory markers IL-6, TNF-α, CRP, COX-2, leptin, SAA1 and IL-8, together with the immunoregulatory cytokine IL-10, were inversely associated with complete involution after age adjustment. Right: single-cell transcriptomic and epigenomic profiling of aged mouse mammary gland reveals coordinated remodeling across stromal, immune and epithelial compartments, with spatial co-localization of aged immune and epithelial cells.

## Figure 6. Proposed model of the menopausal transition as a biological control point for ARLI trajectory

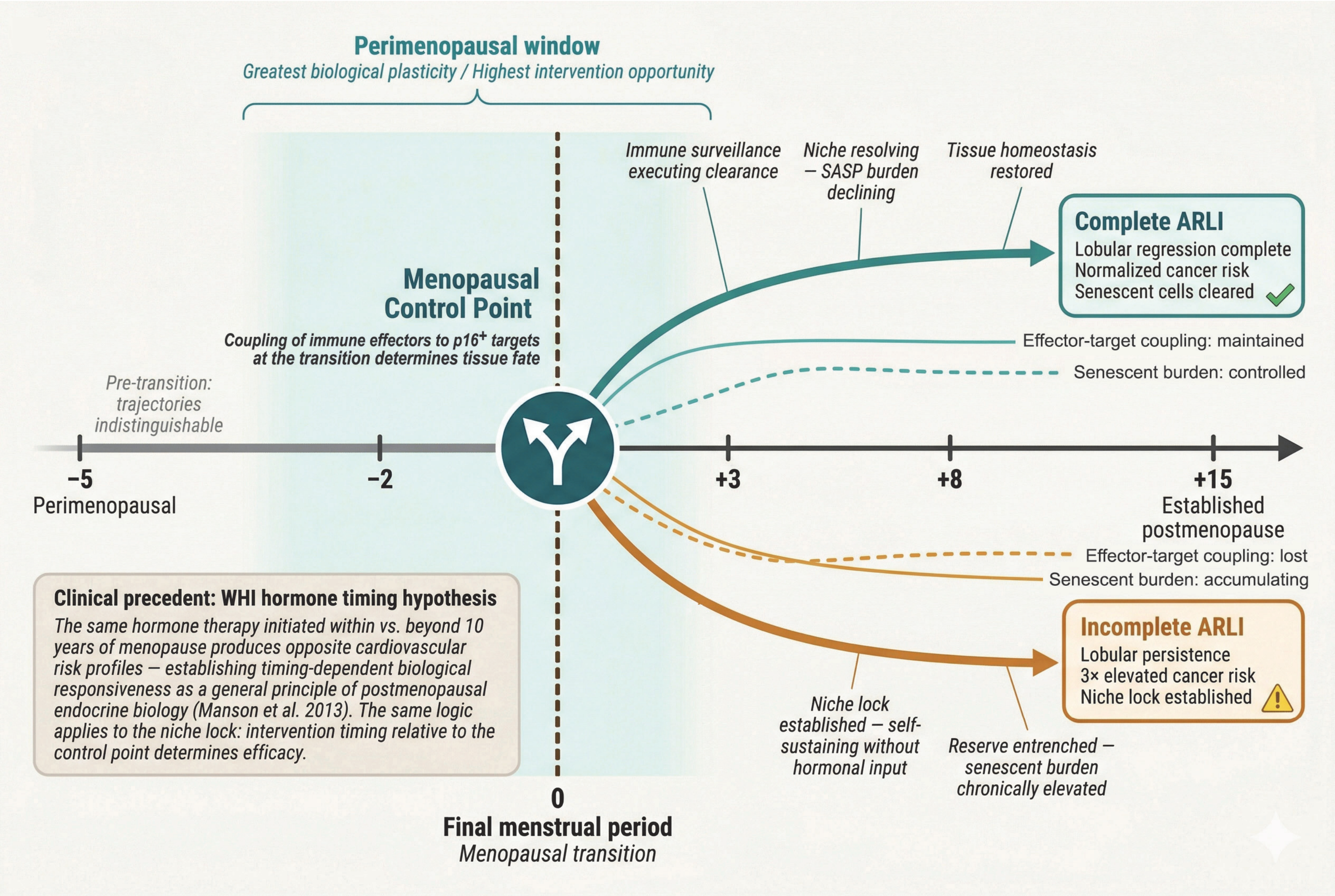

**Figure 6. Proposed model of the menopausal transition as a biological control point for ARLI trajectory**. The x-axis represents years from the final menstrual period, from perimenopause through established postmenopause. Before the transition, tissue fate trajectories are not yet divergent; what is proposed to matter at the control point is the spatial coupling between immune effectors and p16INK4a-positive (senescence-associated) epithelium, together with senescence-associated burden, rather than chronological age alone. Two post-transition trajectories are shown: a resolving trajectory (teal), in which effectors engage and clear senescence-associated cells, SASP burden declines, and lobular regression completes toward lower risk; and a locking trajectory (amber), in which effectors are recruited but are not positioned to engage their targets, senescence-associated burden rises, and a self-sustaining configuration is established. The locking trajectory is not proposed to reflect a loss of immune cells or of immune capacity: in human tissue, noninvoluted postmenopausal lobules carry higher leukocyte densities than completely involuted lobules, while those leukocytes accumulate at the lobular boundary and are depleted from periacinar neighborhoods (Lue et al. 2026). Companion traces depict effector–target coupling and senescence-associated burden along each trajectory; these traces illustrate the proposed model and are not measured quantities. The shaded perimenopausal window represents the period of greatest biological plasticity and therefore the period of greatest intervention opportunity. The timing principle has a clinical precedent: responsiveness to menopausal hormone therapy depends on the state of the target tissue at the time therapy is initiated (Manson et al. 2013). The same logic is proposed to govern reserve niche interventions.

**Figure 7. Comparison of classical view and reserve niche model.**

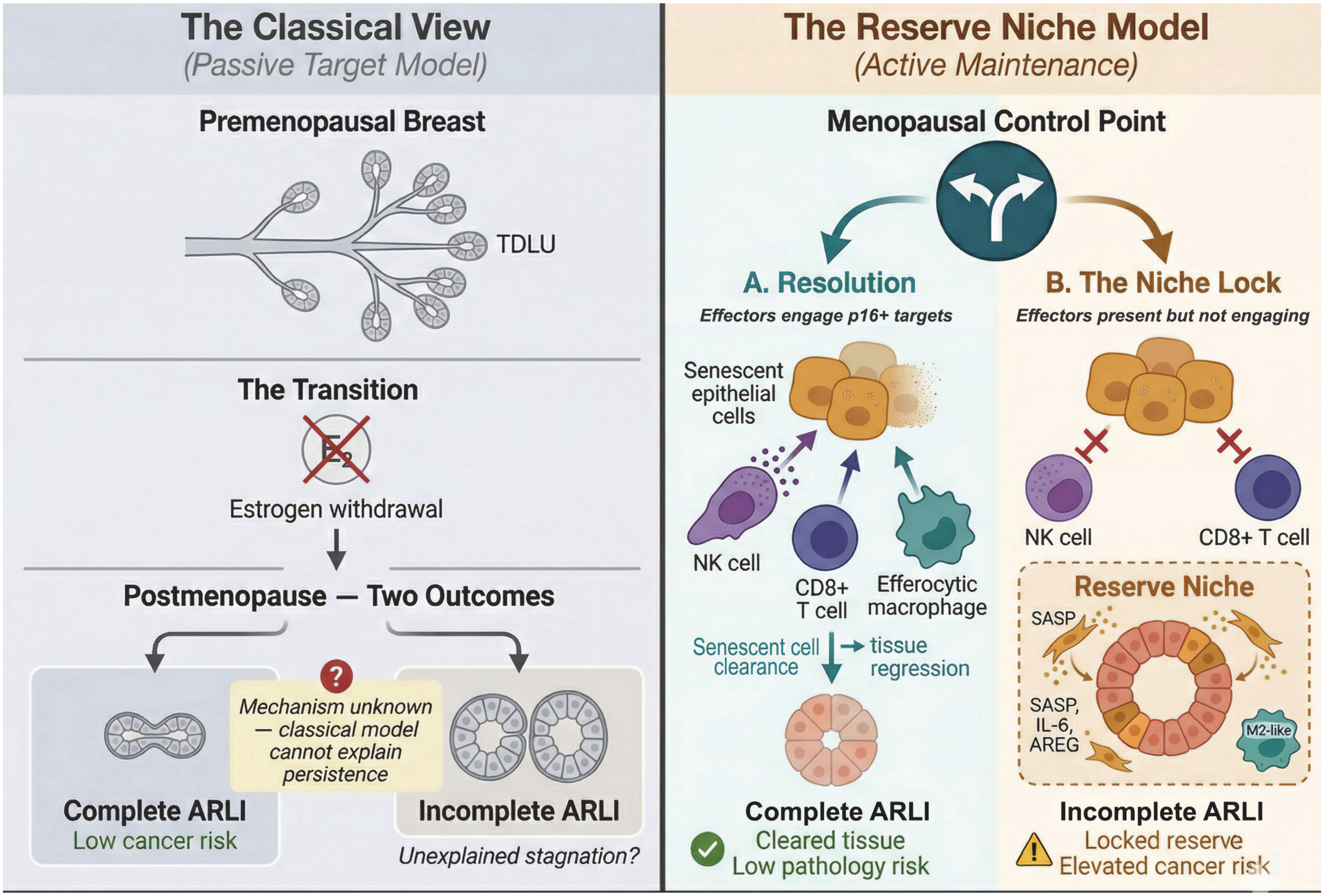

**Figure 7. Comparison of classical view and reserve niche model.** The classical passive target model interprets incomplete age-related lobular involution (ARLI) as residual epithelial tissue available for oncogenic transformation; it accounts for risk in proportion to residual epithelial quantity, but does not by itself explain why involution stalls in some women and not in others. The reserve niche model proposes that the menopausal transition is a biological control point at which the spatial coupling between immune effectors and their targets determines tissue fate. When effectors are positioned to engage p16INK4a-positive (senescence-associated) epithelium, those cells are cleared and lobular regression completes (Resolution, left). When effectors are present but not engaged, senescence-associated cells persist and are proposed to sustain residual epithelium through paracrine signaling, EGFR ligands, and an M2-like macrophage niche (The Niche Lock, right); "niche lock" here denotes spatial uncoupling of surveillance from its targets rather than a loss of immune capacity. An architectural explanation, in which matrix barriers or deficient chemokine gradients prevent effectors from reaching targets they would otherwise engage, would produce the same configuration. This reframes risk stratification from residual tissue volume toward microenvironmental senescence-associated burden and the spatial organization of immune effectors.

## Figure 8. Three mechanistically grounded intervention opportunities mapped onto the post-transition locking trajectory, with timing-dependent efficacy

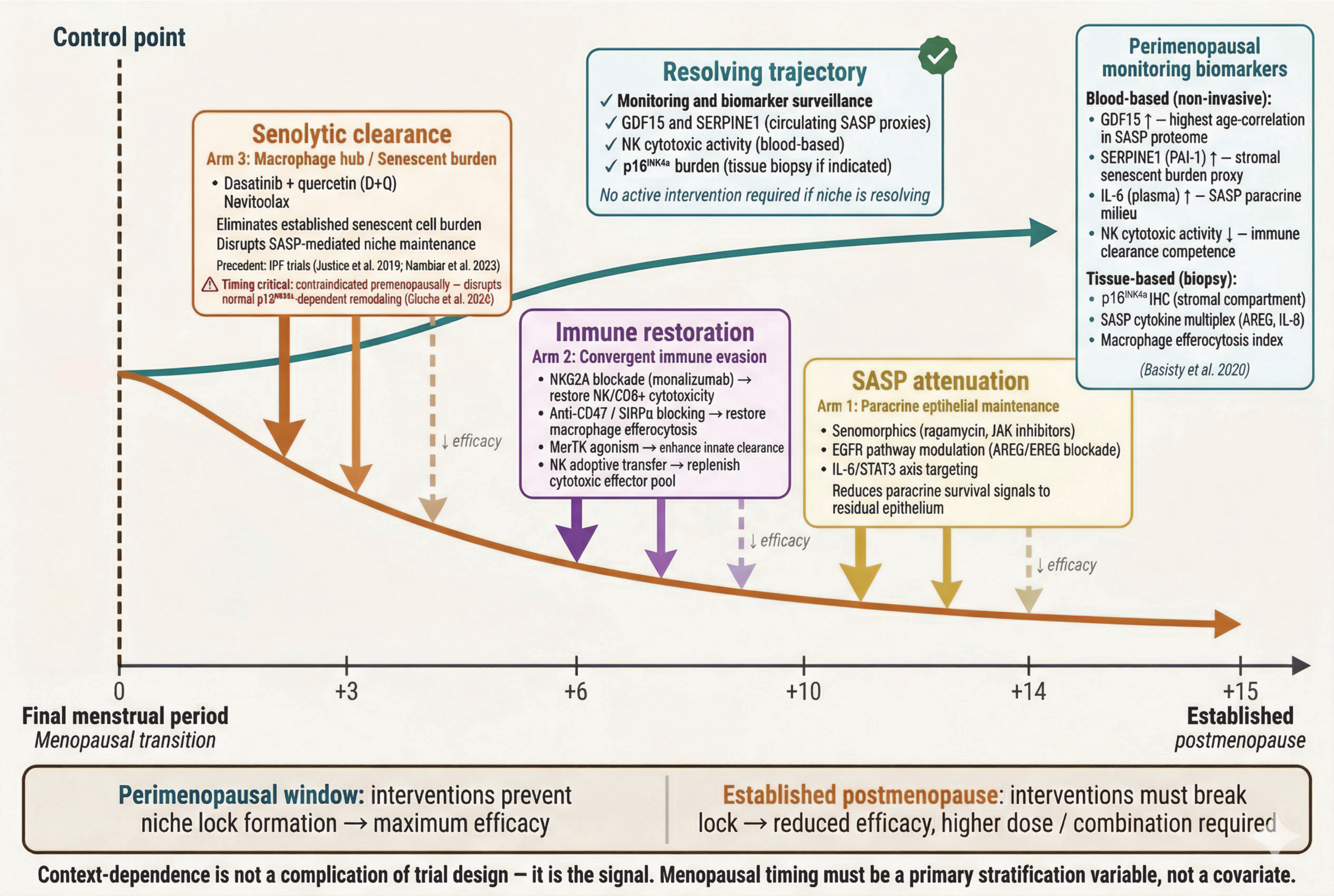

**Figure 8. Proposed mechanism-based intervention map for the senescent-immune niche lock.** The x-axis runs from the menopausal control point to established postmenopause. The resolving trajectory (teal) requires monitoring rather than active intervention. The locking trajectory (amber) presents three intervention classes, each mapped to a mechanistic arm of the hypothesis and each expected to show timing-dependent efficacy. Arrow weight encodes predicted timing-dependent efficacy, with earlier intervention expected to be more effective than later intervention once the niche becomes self-sustaining. Senolytic clearance (Arm 3): dasatinib plus quercetin or navitoclax are proposed to reduce established senescent burden and disrupt SASP-dependent niche maintenance; clinical feasibility has been demonstrated for dasatinib plus quercetin in idiopathic pulmonary fibrosis, whereas navitoclax provides primary senolytic precedent in multiple senescent cell models. Immune restoration (Arm 2): NKG2A blockade, anti-CD47/SIRPα approaches, strategies aimed at restoring MerTK-dependent efferocytosis, and NK adoptive transfer are proposed to address the convergent clearance failures of the niche lock. SASP attenuation (Arm 1): senomorphics, EGFR pathway modulation, and IL-6/STAT3 axis targeting are proposed to reduce paracrine survival signals to residual epithelium. Perimenopausal monitoring biomarkers, shown at upper right, represent candidate indicators for identifying women at risk of niche lock formation during the window of greatest intervention opportunity. The footer encodes the central translational principle: menopausal timing relative to the control point is not merely a covariate but a primary stratification variable that is likely to determine intervention efficacy.